\documentclass[11pt]{article}
\usepackage{acl2016}
\usepackage{times}
\usepackage{latexsym}

\usepackage{amssymb}
\usepackage{amsmath}
\usepackage{gensymb}
\usepackage{stmaryrd}
\usepackage{multirow}
\usepackage{tikz}
\usetikzlibrary{arrows}
\usetikzlibrary{decorations.pathreplacing}
\usetikzlibrary{positioning}
\usepackage{hyperref}
\usepackage{enumitem}
\aclfinalcopy

\newcommand{\myo}[1]{\mathbf{O}_{#1}}

\newcommand{\mygram}{G}

\newcommand{\de}{\leftarrow}
\newcommand{\ep}{\varepsilon}

\newcommand{\dist}{\mathit{d}}
\newcommand{\sub}{\mathit{sub}}
\newcommand{\seg}{\mathit{seg}}
\newcommand{\fromright}{\mathit{from\_right}}
\newcommand{\fromleft}{\mathit{from\_left}}
\newcommand{\toright}{\mathit{to\_right}}
\newcommand{\toleft}{\mathit{to\_left}}

\author{Mark-Jan Nederhof \\
	School of Computer Science\\
	University of St Andrews, UK}

\title{A short proof that $\myo{2}$ is an MCFL}
\date{}

\begin{document}
\maketitle

\begin{abstract} 
We present a new proof that $\myo{2}$ is a multiple context-free language. It contrasts with a recent proof by Salvati (2015) in its avoidance of concepts that seem specific to two-dimensional geometry, such as the complex exponential function. Our simple proof creates realistic prospects of widening the results to higher dimensions. This finding is of central importance to the relation between extreme free word order and classes of grammars used to describe the syntax of natural language.
\end{abstract}

\section{Introduction}

The alphabet of the \emph{MIX language} has three symbols, $a$, $b$ and $c$.
A string is in the language if and only if the number of $a$'s, 
the number of $b$'s, and
the number of $c$'s are all the same. A different way of defining the MIX language is
as permutation closure of the regular language $(\mathit{abc})^*$, as noted by \newcite{BA81a};
see also \newcite{PU83}.

If $a$, $b$ and $c$ represent, say, a transitive verb and its subject and its object, 
then a string in MIX represents a sentence with any number of triples of these constituents,
in a hypothetical language with extreme free word order. This is admittedly rather
unlike any actual natural language. \newcite{JO85} argued that because of this,
grammatical formalisms for describing natural languages should \emph{not} be capable
of generating MIX. He also conjectured that MIX was beyond the generative
capacity of one particular formalism, namely the tree adjoining grammars.
Several decades passed before \newcite{KA12a} finally proved this conjecture.

MIX has been studied in the context of several other formalisms.
\newcite{JO91a} showed that MIX is generated by a generalization of tree adjoining grammars
that decouples local domination for linear precedence.
\newcite{BO99} showed that MIX is generated by a range concatenation grammar.
Negative results were addressed by \newcite{SO14} for well-nested multiple context-free
grammars, and by \newcite{CA09} for a class of categorial grammars.
The MIX language is also of interest outside of computational linguistics,
e.g.\ in computational group theory \cite{GI05}.

A considerable advance in the understanding of the MIX language is due to
\newcite{SA15}, who showed that MIX is generated by a 
multiple context-free grammar (MCFG).
The main part of the proof shows that the language $\myo{2}$ is generated by a MCFG.
This language has four symbols,
$a$, $\overline{a}$,
$b$ and $\overline{b}$. A string is in the language if and only if the number of $a$'s
equals the number of $\overline{a}$'s, and the number of $b$'s
equals the number of $\overline{b}$'s. MIX and $\myo{2}$
are rationally equivalent, which means that if one is generated by a 
multiple context-free grammar, then so is the other.

The proof by \newcite{SA15} is remarkable, in that it is one of the few examples
of geometry being used to prove a statement about formal languages.
The proof has two related disadvantages however. The first is that
a key element of the proof, that of the complex exponential function,
is not immediately understood without background in geometry.
The second is that this also seems to restrict
the proof technique to two dimensions, and there is no obvious avenue to 
generalize the result to a variant of MIX with four or five symbols.
We hope to remedy this by an alternative, self-contained proof
that avoids the complex exponential function. The core idea is a
straightforward normalization of paths in two dimensions, which allow simple
arguments to lead to a proof by contradiction. We also sketch part of a
possible proof in three dimensions.

\section{Initial problem}
\label{sec:problem}

The MCFG $\mygram$ is defined as:
\begin{eqnarray}
S(xy) &\de& R(x,y) \label{rule:s} \\
R(xp,yq) &\de& R(x,y)\ R(p,q) \label{rule:xp,yq} \\
R(xp,qy) &\de& R(x,y)\ R(p,q) \label{rule:xp,qy} \\
R(xpy,q) &\de& R(x,y)\ R(p,q) \label{rule:xpy,q} \\
R(p,xqy) &\de& R(x,y)\ R(p,q) \label{rule:p,xqy} \\
R(a,\overline{a}) &\de&  \label{rule:a} \\
R(\overline{a},a) &\de&  \label{rule:aover} \\
R(b,\overline{b}) &\de&  \label{rule:b} \\
R(\overline{b},b) &\de&  \label{rule:bover} \\
R(\ep,\ep) &\de&  \label{rule:ep}  
\end{eqnarray}
For the meaning of MCFGs in general, 
see \newcite{SE91}; for a closely related formalism,
see \newcite{VI87a}; see \newcite{KA10a} for an overview of 
mildly context-sensitive grammar formalisms.

The reader unfamiliar with this literature is encouraged to interpret the rules
of the grammar as logical implications, with $S$ and $R$ representing
predicates. 
There is an implicit conjunction between
the two occurrences of $R$ in the right-hand side of each of the rules
\mbox{(\ref{rule:xp,yq}) --- (\ref{rule:p,xqy})}.
The symbols $x$, $y$, $p$, $q$ are string-valued
variables, with implicit universal quantification that has scope over both
left-hand side and right-hand side of a rule.
The rules \mbox{(\ref{rule:a}) --- (\ref{rule:ep})} act as axioms. 
The symbols 
$a$, $\overline{a}$, $b$, $\overline{b}$ are terminals, and 
$\ep$ denotes the empty string. 

We can derive $S(x)$ for certain strings $x$, and 
$R(x,y)$ for certain strings $x$ and $y$. 
Figure~\ref{fig:deriv} presents an example of a derivation.
\begin{figure}[t]
\begin{center}
\begin{tikzpicture}[
        level 1/.style={level distance=1.5cm},
        level 2/.style={sibling distance=3.9cm, level distance=1.0cm},
        level 3/.style={sibling distance=2.2cm},
		rule/.style={edge from parent/.style={node distance=5}}
]
\node (n0) {$S(a \overline{b} \overline{a} \overline{b} \overline{a} b b a)$}
 child {node (n1) {$R(a \overline{b} \overline{a} \overline{b} \overline{a} b, b a)$}
  child {node (n2) {$R(a \overline{b}, \overline{a} b)$}
   child {node (n3) {$R(a, \overline{a})$}
   }
   child {node (n4) {$R(\overline{b}, b)$}
   }
  }
  child {node (n5) {$R(\overline{a} \overline{b}, ba)$}
   child {node (n6) {$R(\overline{a}, a)$}
   }
   child {node (n7) {$R(\overline{b}, b)$}
   }
  }
 }
;
\node [below right=-0.2cm and -1cm of n0] {(\ref{rule:s})};
\node [below=-0.2cm of n1] {(\ref{rule:xpy,q})};
\node [below=-0.2cm of n2] {(\ref{rule:xp,yq})};
\node [below=-0.2cm of n5] {(\ref{rule:xp,qy})};
\node [below=-0.2cm of n3] {(\ref{rule:a})};
\node [below=-0.2cm of n4] {(\ref{rule:bover})};
\node [below=-0.2cm of n6] {(\ref{rule:aover})};
\node [below=-0.2cm of n7] {(\ref{rule:bover})};
\end{tikzpicture}
\end{center}
\caption{Derivation in $\mygram$. The numbers indicate the rules that were
used.}
\label{fig:deriv}
\end{figure}
The language generated by $\mygram$ is the set $L$ of strings $x$ such that $S(x)$
can be derived.

By induction on the depth of derivations, 
one can show that if $R(x,y)$, for strings $x$ and $y$,
then $xy \in \myo{2}$.
Thereby, if $S(x)$ then $x\in\myo{2}$, which means $L\subseteq \myo{2}$.
The task ahead is to prove that 
if $xy \in \myo{2}$, for some $x$ and $y$, then $R(x,y)$. From this, $L= \myo{2}$ then follows.

Let $|x|$ denote the length of string $x$.
For an inductive proof that $xy \in \myo{2}$ implies $R(x,y)$, the base cases are as follows.
If $xy \in \myo{2}$ and $|x|\leq 1$ and $|y|\leq 1$, then
trivially $R(x,y)$ by rules \mbox{(\ref{rule:a}) --- (\ref{rule:ep})}.

Furthermore, if we can prove that $xy \in \myo{2}$,
$x\neq\ep$ and $y\neq\ep$ together imply $R(x,y)$, for $|xy|=m$, some $m$,
then we may also prove that $x'y' \in \myo{2}$ on its own implies $R(x',y')$ for $|x'y'|=m$.
To see this, consider $m>0$ and $z \in \myo{2}$ with $|z| = m$,
and write it as $z=xy$ for some $x\neq\ep$ and $y\neq\ep$.
If by assumption $R(x,y)$, then together with $R(\ep,\ep)$ and
rule (\ref{rule:xpy,q}) or (\ref{rule:p,xqy}) we may derive $R(xy,\ep)$ or
$R(\ep,xy)$,
respectively.
In the light of this, the inductive step merely needs to show
that if for some $x$ and $y$:
\begin{itemize}
\item $xy \in \myo{2}$, $|x| \geq 1$, $|y| \geq 1$ and $|xy| > 2$, and
\item $pq\in \myo{2}$ and $|pq|<|xy|$ imply $R(p,q)$, for all $p$ and $q$, 
\end{itemize}
then also $R(x,y)$. 
One easy case is if $x\in \myo{2}$
(and thereby $y\in \myo{2}$)
because then we can write $x=x_1x_2$ for some
$x_1\neq\ep$ and $x_2\neq\ep$.
The inductive hypothesis states that $R(x_1,x_2)$ and $R(\ep, y)$,
which imply $R(x,y)$ using rule (\ref{rule:xpy,q}).

A second easy case is if $x$ or $y$ has a proper prefix or proper suffix 
that is in $\myo{2}$. For example,
assume there are $z_1\neq\ep$ and $z_2\neq\ep$ such that $x=z_1 z_2$ and
$z_1\in \myo{2}$.
Then we can use the inductive hypothesis on $R(z_1,\ep)$ and $R(z_2,y)$,
together with rule (\ref{rule:xp,yq}).

At this time, the reader may wish to read Figure~\ref{fig:deriv} from the root
downward. First, $a \overline{b} \overline{a} \overline{b}
\overline{a} b b a$ is divided into a pair of strings, namely
$a \overline{b} \overline{a} \overline{b} \overline{a} b$ and $b a$.
At each branching node in the derivation, a pair of strings is divided into
four strings, which are grouped into two pairs of strings, using 
rules \mbox{(\ref{rule:xp,yq}) --- (\ref{rule:p,xqy})}, read from left to
right. Rules (\ref{rule:xp,yq}) and (\ref{rule:xp,qy}) divide
each left-hand side argument into two parts. Rule (\ref{rule:xpy,q})
divides the first left-hand side argument into three parts, and rule
(\ref{rule:p,xqy}) divides the second left-hand side argument into
three parts. 

What remains to show is that if
$z_1z_2 \in \myo{2}$, $z_1\notin \myo{2}$ and $|z_1z_2| > 2$, 
and no proper prefix or proper suffix of $z_1$ or of $z_2$ is in $\myo{2}$,
then there is at least one rule that allows us to
divide $z_1$ and $z_2$ into four strings altogether,
say $x,y,p,q$, of which at least three are non-empty, such that
$xy \in \myo{2}$. This will then permit use of the inductive hypothesis
on $R(x,y)$ and on $R(p,q)$.

We can in fact restrict our attention to 
$z'_1z'_2 \in \myo{2}$, $|z'_1z'_2| > 2$,
and no non-empty substring of $z'_1$ or of $z'_2$ is in
$\myo{2}$, which can be justified as follows. Suppose we have $z_1$ and $z_2$ as
in the previous paragraph, and suppose $z'_1$ and $z'_2$ result from $z_1$ and $z_2$
by exhaustively removing all non-empty substrings that are in $\myo{2}$; note that
still $|z'_1z'_2| > 2$. If we can use a rule to divide $z'_1$ and $z'_2$ into 
$x',y',p',q'$, of which at least three are non-empty, such that
$x'y' \in \myo{2}$, then the same rule can be used to divide $z_1$ and $z_2$ into
$x,y,p,q$ with the required properties, which can be 
found from $x',y',p',q'$ 
simply by reintroducing the removed substrings at corresponding positions.

\section{Geometrical view}
\label{sec:geo}

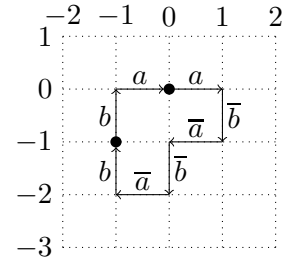
\begin{figure}[t]
\begin{center}
\begin{tikzpicture}[scale=0.70,
astyle/.style = {above,yshift=-2.0pt},
bstyle/.style = {below,yshift=2.0pt},
rstyle/.style = {right,xshift=-2.0pt},
lstyle/.style = {left,xshift=2.0pt},
]
\draw[thin,dotted] (-2,-3) grid (2,1);
\draw [->] (0,0) -- node[astyle] {$a$} (1,0);
\draw [->] (1,0) -- node[rstyle] {$\overline{b}$} (1,-1);
\draw [->] (1,-1) -- node[astyle] {$\overline{a}$} (0,-1);
\draw [->] (0,-1) -- node[rstyle] {$\overline{b}$} (0,-2);
\draw [->] (0,-2) -- node[astyle] {$\overline{a}$} (-1,-2);
\draw [->] (-1,-2) -- node[lstyle] {$b$} (-1,-1.1);
\draw [->] (-1,-1) -- node[lstyle] {$b$} (-1,0);
\draw [->] (-1,0) -- node[astyle] {$a$} (-0.1,0);
\fill[black] (0, 0) circle[radius=3pt];
\fill[black] (-1, -1) circle[radius=3pt];
\node[above,xshift=-0.0cm,yshift=-0.0cm] at (-2,1) {$-2$};
\node[above,xshift=-0.0cm,yshift=-0.0cm] at (-1,1) {$-1$};
\node[above,xshift=-0.0cm,yshift=-0.0cm] at (0,1) {$0$};
\node[above,xshift=-0.0cm,yshift=-0.0cm] at (1,1) {$1$};
\node[above,xshift=-0.0cm,yshift=-0.0cm] at (2,1) {$2$};
\node[left,xshift=0.0pt,yshift=-0.0cm] at (-2,1) {$1$};
\node[left,xshift=0.0pt,yshift=-0.0cm] at (-2,0) {$0$};
\node[left,xshift=0.0pt,yshift=-0.0cm] at (-2,-1) {$-1$};
\node[left,xshift=0.0pt,yshift=-0.0cm] at (-2,-2) {$-2$};
\node[left,xshift=0.0pt,yshift=-0.0cm] at (-2,-3) {$-3$};
\end{tikzpicture}
\end{center}
\caption{Two strings $x=a \overline{b} \overline{a} \overline{b}
\overline{a} b$ and $y=\mathit{b a}$ together represent a closed path, consisting
of a path from $(0,0)$ to $(-1,-1)$ and a path from $(-1,-1)$ to $(0,0)$.}
\label{fig:path}
\end{figure}

We may interpret a string $x$ geometrically
in two dimensions, as a path consisting of a series
of line segments of length 1, starting in some point $(i,j)$.
Every symbol
in $x$, from beginning to end, represents the next line segment in
that path; an occurrence of
$a$ represents a line segment from the previous point $(i,j)$ to the next point $(i+1,j)$,
$\overline{a}$ represents a line segment from $(i,j)$ to $(i-1,j)$,
$b$ represents a line segment from $(i,j)$ to $(i,j+1)$,
and $\overline{b}$ represents a line segment from $(i,j)$ to $(i,j-1)$.
If $x\in\myo{2}$, then the path is \emph{closed}, that is, the starting
point and the ending point are the same.
If we have two strings $x$ and $y$
such that $xy\in\myo{2}$ and $x \notin \myo{2}$, then this translates
to two paths, connecting two distinct points, which together form a closed path.
This is illustrated in Figure~\ref{fig:path}.

In the following, we assume a fixed choice of some $x$ and $y$ such that
$xy\in\myo{2}$, 
$|xy| > 2$, 
and no non-empty substring 
of $x$ or of $y$ is in $\myo{2}$.
The path of $x$
starting in $(0,0)$ will be called $A[0]$. Let $(i,j)$ be the point where this path
ends; $i$ is the number of occurrences of $a$ in $x$ minus the number of occurrences
of $\overline{a}$ in $x$; $j$ is similarly determined. The path of
$y$, from $(i,j)$ back to $(0,0)$, will be called $B[1]$. We generalize this
to paths called $A[k]$ and $B[k]$, for integer $k$, which are the paths of
$x$ and $y$, respectively, that start in $(k\cdot i,k\cdot j)$.
Where the starting points are irrelevant, we talk about paths $A$ and $B$.
We also refer to $(k\cdot i,k\cdot j)$ as point $P[k]$.

Let $C$ be a path, which can be either $A[k]$ or $B[k]$ for some $k$.
We write $Q\in C$ to denote that $Q$ is a point on $C$.
Let $Q=(i,j)\in C$, not necessarily with $i$ and $j$ being integers.
We define the path-distance $\dist_{C}(Q)$ of
$Q$ on $C$ to be the length of the 
path along line segments of $C$
to get from $P[k]$ to $Q$.  
In Figure~\ref{fig:path},
$(0,-1)$ has path-distance 3 on $A[0]$, as the 
path on $A[0]$ 
to reach $(0,-1)$ from $P[0]=(0,0)$ consists of 
the line segments represented by the prefix 
$a \overline{b} \overline{a}$ of $x$. Similarly, $\dist_{A[0]}((0.5,-1)) = 2.5$.


Let $C$ be a path as above and let points $Q_1,Q_2\in C$ be such that $\dist_C(Q_1) \leq \dist_C(Q_2)$.
We define the subpath $D$ = $\sub_C(Q_1,Q_2)$ to be
such that $Q\in D$ if and only if $Q\in C$ and $\dist_C(Q_1) \leq Q \leq \dist_C(Q_2)$, and
$\dist_D(Q) = \dist_C(Q)-\dist_C(Q_1)$ for every $Q\in D$.
For two points $Q_1$ and $Q_2$, the line segment between $Q_1$ and $Q_2$ is
denoted by $\seg(Q_1,Q_2)$.

The task formulated at the end of Section~\ref{sec:problem} is accomplished
if we can show that at least one of the following must hold:
\begin{itemize}
\item the angle in $P[0]$ between the beginning of $A[0]$ and that of $B[0]$ is $180\degree$ 
(Figure~\ref{fig:A0andB0});
\item there is a point $Q\notin\{P[0],P[1]\}$ such that $Q\in A[0]$ and $Q\in B[1]$
(Figure~\ref{fig:A0andB1});
\item there is a point $Q\neq P[1]$ such that $Q\in A[0]$, $Q\in A[1]$ and $\dist_{A[0]}(Q) > \dist_{A[1]}(Q)$
(Figure~\ref{fig:A0andA1}); or
\item there is a point $Q\neq P[0]$ such that $Q\in B[0]$, $Q\in B[1]$ and $\dist_{B[1]}(Q) > \dist_{B[0]}(Q)$
(analogous to Figure~\ref{fig:A0andA1}).
\end{itemize}
\begin{figure}[t]
\begin{center}
\begin{tikzpicture}[scale=0.70,
astyle/.style = {above,yshift=-2.0pt},
bstyle/.style = {below,yshift=2.0pt},
rstyle/.style = {right,xshift=-2.0pt},
lstyle/.style = {left,xshift=2.0pt},
]
\draw [->] (0,0) -- node[astyle] {$a$} (1,0);
\draw [->] (1,0) -- node[rstyle] {$b$} (1,1);
\draw [->] (1,1) -- node[astyle] {$a$} (2,1);
\draw [->] (2,1) -- node[rstyle] {$\overline{b}$} (2,0);
\draw [->] (2,0) -- node[rstyle] {$\overline{b}$} (2,-0.9);
\draw [->] (2,-1) -- node[bstyle] {$\overline{a}$} (1,-1);
\draw [->] (1,-1) -- node[bstyle] {$\overline{a}$} (0,-1);
\draw [->] (0,-1) -- node[rstyle] {$b$} (0,-0.1);
\draw [->,dotted] (0,0) -- node[astyle] {$\overline{a}$} (-1,0);
\draw [->,dotted] (-1,0) -- node[astyle] {$\overline{a}$} (-2,0);
\draw [->,dotted] (-2,0) -- node[lstyle] {$b$} (-2,0.9);
\fill[black] (0, 0) circle[radius=3pt];
\fill[black] (2, -1) circle[radius=3pt];
\fill[black] (-2, 1) circle[radius=3pt];
\node[below left,font=\small,xshift=0pt,yshift=0pt] at (0,0) {$P[0]$};
\node[below right,font=\small,xshift=-2pt,yshift=2pt] at (2,-1) {$P[1]$};
\node[above left,font=\small,xshift=0pt,yshift=-3pt] at (-2,1) {$P[-1]$};
\draw[->, dashed] (-0.9,0.2) arc (170:10:0.9);
\node[above,font=\small,xshift=0pt,yshift=0pt] at (0,1) {$180\degree$};
\end{tikzpicture}
\end{center}
\caption{With $x=a\,b\,a\,\overline{b}\,\overline{b}$
and $y=\overline{a}\,\overline{a}\,b$,
the beginning of path $A[0]$ and the beginning of (dotted) path $B[0]$ have an $180\degree$ angle in $P[0]$,
which implies $x$ and $y$ start with complementing symbols (here $a$ and $\overline{a}$;
the other possibility is $b$ and $\overline{b}$). 
By applying rule~(\ref{rule:xp,yq}), two smaller closed paths result,
one of which consists of these two complementing symbols.}
\label{fig:A0andB0}
\end{figure}
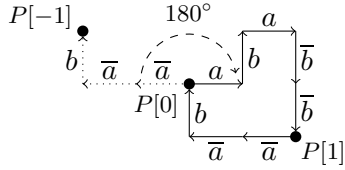
\begin{figure}[t]
\begin{center}
\begin{tikzpicture}[scale=0.70,
astyle/.style = {above,yshift=-2.0pt},
bstyle/.style = {below,yshift=2.0pt},
rstyle/.style = {right,xshift=-2.0pt},
lstyle/.style = {left,xshift=2.0pt},
]
\draw [->] (0,0) -- node[lstyle] {$b$} (0,1);
\draw [->] (0,1) -- node[astyle] {$a$} (0.9,1);
\draw [->] (1.1,1) -- node[astyle] {$a$} (2,1);
\draw [->] (2,1) -- node[astyle] {$a$} (3,1);
\draw [->] (3,1) -- node[rstyle] {$b$} (3,1.9);
\draw [->] (3,2) -- node[astyle] {$\overline{a}$} (2,2);
\draw [->] (2,2) -- node[astyle] {$\overline{a}$} (1,2);
\draw [->] (1,2) -- node[rstyle] {$\overline{b}$} (1,1.1);
\draw [->] (1,0.9) -- node[rstyle] {$\overline{b}$} (1,0);
\draw [->] (1,0) -- node[astyle] {$\overline{a}$} (0.1,0);
\fill[black] (0, 0) circle[radius=3pt];
\fill[black] (3, 2) circle[radius=3pt];
\draw[black] (1, 1) circle[radius=3pt];
\node[below left,font=\small,xshift=2pt,yshift=2pt] at (0,0) {$P[0]$};
\node[above right,font=\small,xshift=-2pt,yshift=-2pt] at (3,2) {$P[1]$};
\end{tikzpicture}
\end{center}
\caption{The paths $A[0]$ and $B[1]$ of $x=\mathit{b a a a b}$ 
and $y=\overline{a}\,\overline{a}\overline{b}\,\overline{b}\overline{a}$
have point $(1,1)$ in common. Two smaller closed paths result by
applying rule~(\ref{rule:xp,qy}).}
\label{fig:A0andB1}
\end{figure}
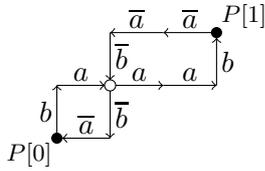
\begin{figure}[t]
\begin{center}
\begin{tikzpicture}[scale=0.70,
astyle/.style = {above,yshift=-2.0pt},
bstyle/.style = {below,yshift=2.0pt},
rstyle/.style = {right,xshift=-2.0pt},
lstyle/.style = {left,xshift=2.0pt},
]
\draw [->] (0,0) -- node[astyle] {$\overline{a}$} (-1,0);
\draw [->] (-1,0) -- node[astyle] {$\overline{a}$} (-2,0);
\draw [->] (-2,0) -- node[astyle] {$\overline{a}$} (-2.9,0);
\draw [->] (-3,0) -- node[lstyle] {$b$} (-3,1);
\draw [->] (-3,1) -- node[lstyle] {$b$} (-3,2);
\draw [->] (-3,2) -- node[astyle] {$a$} (-2,2);
\draw [->] (-2,2) -- node[astyle] {$a$} (-1,2);
\draw [->] (-1,2) -- node[lstyle] {$\overline{b}$} (-1,1);
\draw [->] (-1,1) -- node[bstyle] {$a$} (-0.1,1);
\draw [->] (0.1,1) -- node[bstyle] {$a$} (1,1);
\draw [->] (1,1) -- node[rstyle] {$\overline{b}$} (1,0);
\draw [->] (1,0) -- node[astyle] {$\overline{a}$} (0.1,0);
\draw [->,dotted] (0,0) -- node[lstyle] {$$} (0,0.9);
\draw [->,dotted] (0,1.1) -- node[lstyle] {$$} (0,2);
\draw [->,dotted] (0,2) -- node[astyle] {$$} (1,2);
\draw [->,dotted] (1,2) -- node[astyle] {$$} (2,2);
\draw [->,dotted] (2,2) -- node[lstyle] {$$} (2,1);
\draw [->,dotted] (2,1) -- node[astyle] {$$} (3,1);
\draw [->,dotted] (3,1) -- node[astyle] {$$} (4,1);
\draw [->,dotted] (4,1) -- node[rstyle] {$$} (4,0);
\draw [->,dotted] (4,0) -- node[astyle] {$$} (3.1,0);
\fill[black] (0, 0) circle[radius=3pt];
\fill[black] (-3, 0) circle[radius=3pt];
\fill[black] (3, 0) circle[radius=3pt];
\draw[black] (0, 1) circle[radius=3pt];
\node[below,font=\small,xshift=0pt,yshift=1pt] at (0,0) {$P[1]$};
\node[below,font=\small,xshift=0pt,yshift=1pt] at (-3,0) {$P[0]$};
\node[below,font=\small,xshift=0pt,yshift=1pt] at (3,0) {$P[2]$};
\node[above right,font=\small,xshift=-2pt,yshift=-2pt] at (0,1) {$Q$};
\end{tikzpicture}
\end{center}
\caption{With $x=b\,b\,a\,a\,\overline{b}\,a\,a\,\overline{b}\,\overline{a}$
and $y=\overline{a}\,\overline{a}\,\overline{a}$,
the path $A[0]$ and the (dotted) path $A[1]$ 
have point $Q$ in common, with $\dist_{A[0]}(Q)=6 > \dist_{A[1]}(Q)=1$.
By applying rule~(\ref{rule:xpy,q}), two smaller closed paths result,
one of which is formed by prefix $b$ of length 1 and suffix
$a\overline{b}\,\overline{a}$ of length $|x|-6=3$ of $x$.}
\label{fig:A0andA1}
\end{figure}
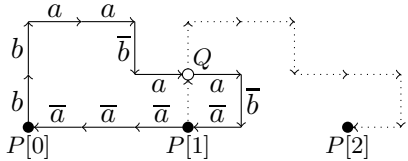

We will do this through a contradiction that results if we assume:
\begin{enumerate}[label=(\roman*)]
\item the angle in $P[0]$ between the beginning of $A[0]$ and that of $B[0]$ is \emph{not} $180\degree$;
\item $A[0]\cap B[1] = \{P[0],P[1]\}$;
\item there is no $Q\in (A[0] \cap A[1])\setminus \{P[1]\}$ such that $\dist_{A[0]}(Q) > \dist_{A[1]}(Q)$; and
\item there is no $Q\in (B[0]\cap B[1])\setminus \{P[0]\}$ such that $\dist_{B[1]}(Q) > \dist_{B[0]}(Q)$.
\end{enumerate}
In the below, we will refer to these assumptions as the \emph{four constraints}.

\section{Continuous view}
\label{sec:continuous}

Whereas paths $A$ and $B$ were initially formed out of line segments of length
1 between
points $(i,j)$ with integers $i$ and $j$, the proof becomes considerably
easier if we allow $i$ and $j$ to be real numbers. The benefit
lies in being able to make changes to the paths that preserve the
four constraints, to obtain a convenient normal form for $A$ and $B$. 
If we can prove a contradiction on the normal form, we will have shown that no
$A$ and $B$ can exist that satisfy the four constraints.

We define, for each integer $k$, the line $\ell[k]$, which is
perpendicular to the line through $P[k]$ and $P[k+1]$, and lies exactly
half-way between $P[k]$ and $P[k+1]$.
Much as before, we write $Q\in \ell[k]$ to denote that $Q$ is a point on line
$\ell[k]$.
We will consistently draw points \ldots, $P[-1]$, $P[0]$, $P[1]$, \ldots in a
straight line from left to right.

Let $C$ be a path, which can be either $A[k']$ or $B[k']$, some $k'$, and
let $Q\in C$.
We write $\fromright_C(Q,\ell[k])$ to mean that path $C$ is strictly to the
right of $\ell[k]$ just before reaching $Q$, or formally,
there is some $\delta>0$ such that each $Q'\in C$ with
$\dist_{C}(Q)-\delta < \dist_{C}(Q') <\dist_{C}(Q)$
lies strictly to the right of $\ell[k]$.
The predicates $\fromleft$, $\toright$, $\toleft$ are similarly defined.

Let $Q_1,Q_2 \in C \cap \ell[k]$, some $k$, such that $\dist_{C}(Q_1) \leq
\dist_{C}(Q_2)$.
We say that $C$ has an \emph{excursion from the right} between $Q_1$ and $Q_2$ at $\ell[k]$ 
if $\fromright_C(Q_1,\ell[k])$ and $\toright_C(Q_2,\ell[k])$.
This is illustrated in
Figure~\ref{fig:excursion}: the path is strictly to the right of $\ell[k]$
just before reaching $Q_1$. From there on it may (but need not) cross over to the left of
$\ell[k]$. Just after it reaches
$Q_2$, it must again be strictly to the right of $\ell[k]$. 
The definition of \emph{excursion from the left} is symmetric.
Note that excursions may be nested; in
Figure~\ref{fig:excursion}, $\sub_C(Q_1,Q_2)$  has an excursion at $\ell[k]$ from the left below $Q_2$.

\begin{figure}[t]
\begin{center}
\begin{tikzpicture}[scale=0.70]
\draw[thick] (1,3.3) -- (0,2.3) -- (0,1.8) -- (-1,1.0) -- 
(-1,-0.25) -- (1.0,-0.25) -- (-0.5,0.75) --
(0,1.0) -- (1.5,1.75);
\draw[-triangle 90] (0.8,3.1) -- +(-0.1,-0.1);
\draw[-triangle 90] (1,1.5) -- +(0.2,0.1);
\draw[dotted] (-3.0,-1) -- (-3,3);
\draw[dotted] (0,-1) -- (0,3);
\draw[dotted] (3,-1) -- (3,3);
\node[above,xshift=0pt] at (-3,3) {$\ell[k-1]$};
\node[above,xshift=0pt] at (0,3) {$\ell[k]$};
\node[above,xshift=0pt] at (3,3) {$\ell[k+1]$};
\draw[fill] (0,2.3) circle [radius=2.0pt];
\draw[fill] (0,1.8) circle [radius=2.0pt];
\draw[fill] (0,1.0) circle [radius=2.0pt];
\node[left,xshift=1pt,yshift=2pt] at (0,2.3) {$Q_1$};
\node[right,xshift=-2pt] at (0,1.8) {$R_1$};
\node[below right,xshift=-3pt,yshift=3pt] at (0,1.0) {$R_2{=}Q_2$};
\begin{scope}[shift={(0,-0.5)}]
\draw[fill] (-1.5,0) circle [radius=2.0pt];
\draw[fill] (1.5,0) circle [radius=2.0pt];
\draw[fill] (4.5,0) circle [radius=2.0pt];
\node[below,xshift=0pt,font=\small] at (-1.5,0) {$P[k]$};
\node[below,xshift=0pt,font=\small] at (1.5,0) {$P[k+1]$};
\node[below,xshift=0pt,font=\small] at (4.5,0) {$P[k+2]$};
\end{scope}
\end{tikzpicture}
\end{center}
\caption{Excursion from the right at $\ell[k]$.}
\label{fig:excursion}
\vspace{1em}
\begin{center}
\begin{tikzpicture}[scale=0.70]
\draw[dashed] (0.5,2.8) -- (0,2.3) -- (0,1.8) -- (-1,1.0) -- 
(-1,-0.25) -- (1.0,-0.25) -- (-0.5,0.75) --
(0,1.0) -- (0.5,1.25);
\draw[-triangle 90] (0.8,3.1) -- +(-0.1,-0.1);
\draw[-triangle 90] (1,1.5) -- +(0.2,0.1);
\draw[thick] (1,3.3) -- (0.5,2.8) -- (0.5,1.25) -- (1.5,1.75);
\draw[dotted] (-3.0,-1) -- (-3,3);
\draw[dotted] (0,-1) -- (0,3);
\draw[dotted] (3,-1) -- (3,3);
\draw[dotted] (0.5,-0.65) -- (0.5,3.5);
\node[above,xshift=0pt] at (-3,3) {$\ell[k-1]$};
\node[above,xshift=0pt] at (0,3) {$\ell[k]$};
\node[above,xshift=0pt] at (3,3) {$\ell[k+1]$};
\node[below,xshift=0pt] at (0.5,-0.65) {$m$};
\draw[fill] (0.5,2.8) circle [radius=2.0pt];
\draw[fill] (0.5,1.25) circle [radius=2.0pt];
\node[below right,xshift=-2pt,yshift=3pt] at (0.5,2.8) {$Q'_1$};
\node[below right,xshift=-2pt,yshift=3pt] at (0.5,1.25) {$Q'_2$};
\begin{scope}[shift={(0,-0.5)}]
\draw[fill] (-1.5,0) circle [radius=2.0pt];
\draw[fill] (1.5,0) circle [radius=2.0pt];
\draw[fill] (4.5,0) circle [radius=2.0pt];
\node[below,xshift=0pt,font=\small] at (-1.5,0) {$P[k]$};
\node[below,xshift=5pt,font=\small] at (1.5,0) {$P[k+1]$};
\node[below,xshift=0pt,font=\small] at (4.5,0) {$P[k+2]$};
\end{scope}
\end{tikzpicture}
\end{center}
\caption{The excursion from Figure~\ref{fig:excursion} truncated in 
$Q'_1$ and $Q'_2$ on line $m$.}
\label{fig:truncation}
\end{figure}

In Figure~\ref{fig:excursion}, 
the pair of points $Q_1$ and $R_1$ will be called a \emph{crossing} of $\ell[k]$ from right
to left, characterized by $Q_1,R_1\in \ell[k]$, 
$\fromright_C(Q_1,\ell[k])$, $\toleft_C(R_1,\ell[k])$
and $\sub_C(Q_1,R_1)$ being a line segment. The pair of points $R_2$ and $Q_2$
is a crossing of $\ell[k]$ from left to right, where the length of
$\seg(R_2,Q_2)$ happens to be 0.
In much of the following we will simplify the discussion
by assuming crossings consist of single points, as in the case of $R_2=Q_2$. 
However, existence of crossings consisting of line segments of non-zero length, 
as in the case of $Q_1$ and $R_1$, would not invalidate any of the arguments of the proof.

Excursions are the core obstacle that needs to be overcome for our proof.
We can \emph{truncate} an excursion at $\ell[k]$ 
by finding a suitable line $m$ that is parallel to $\ell[k]$, some small distance
away from it, between $\ell[k]$
and $P[k+1]$ for excursions from the right, and between $\ell[k]$ and $P[k]$ 
for excursions from the left.
We further need to find points $Q'_1, Q'_2 \in C \cap m$, where
$\dist_C(Q'_1) < \dist_C(Q_1)$ and $\dist_C(Q_2) < \dist_C(Q'_2)$. 
Because our coordinates no longer need to
consist of integers, it is clear that
$m$, $Q'_1$ and $Q'_2$ satisfying these requirements must exist.

The truncation consists in changing $\sub_C(Q'_1,Q'_2)$ to become
$\seg(Q'_1,Q'_2)$, as illustrated by
Figure~\ref{fig:truncation}. Note that if $C$ is say $A[k']$, some $k'$, then changing
the shape of $C$ means changing the shape of $A[k'']$ for any other $k''$ as well;
the difference between $A[k']$ and $A[k'']$ is only in the starting point
$P[k']$ versus $P[k'']$.

At this time, we must allow for the possibility
that for some excursions, no $m$, $Q'_1$ and $Q'_2$
can be found with which we can implement a truncation, if we also need to 
preserve the four constraints and preserve absence of self-intersections.
There is a small number of possible causes.
First, suppose that $C=A[k']$ and $B[k'+1]$ intersects with
$\seg(Q_1,Q_2)$. Then $B[k'+1]$ may intersect with $\seg(Q'_1,Q'_2)$
for any choice of $m$, $Q'_1$ and $Q'_2$, and thereby
no truncation is possible without violating constraint~(ii).
Similarly, a truncation may be blocked if $C=B[k'+1]$ and $A[k']$ intersects with $\seg(Q_1,Q_2)$.
Next, it could be that
$C=A[k']$, while 
$\dist_{A[k']}(Q_1) > \dist_{A[k'+1]}(Q)$ holds for some
$Q \in \seg(Q_1,Q_2)\cap A[k'+1]$, or
$\dist_{A[k'-1]}(Q) > \dist_{A[k']}(Q_2)$ holds for some
$Q\in\seg(Q_1,Q_2)\cap A[k'-1]$, either of which potentially blocks
a truncation if constraint~(iii) is to be preserved.
Constraint~(iv) has similar consequences.
Furthermore, if we need to preserve absence of self-intersections, 
a truncation may be blocked if $\dist_C(Q) < \dist_C(Q_1)$ or $\dist_C(Q_2) < \dist_C(Q)$
for some $Q\in \seg(Q_1,Q_2)\cap C$.

\section{Normal form}

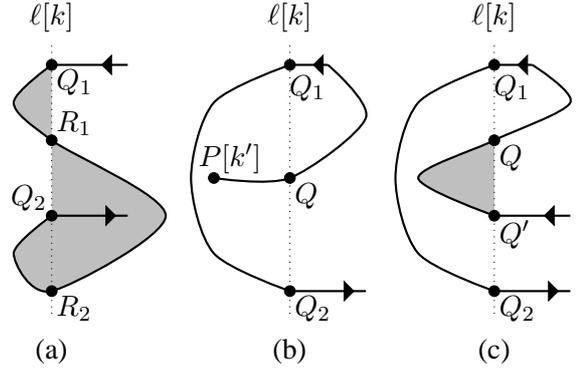
\begin{figure}[t]
\begin{center}
\begin{tikzpicture}
\begin{scope}
\clip plot [smooth] coordinates {(0,2.5) (-0.5,2.0) (0,1.5) (1.5,0.5)
(0,-0.5) (-0.5,-0.0) (0,0.5)};
\fill[color=lightgray] (-2,-2) rectangle (5,5);
\end{scope}
\draw[dotted] (0,-0.8) -- (0,2.8);
\node[above,xshift=0pt] at (0,2.8) {$\ell[k]$};
\node[below right,xshift=-2pt,yshift=2pt] at (0,2.5) {$Q_1$};
\draw[thick] (1,2.5) -- (0,2.5);
\draw[fill] (0,2.5) circle [radius=2.0pt];
\draw[-triangle 90] (0.75,2.5) -- +(-0.1,-0.0);
\node[above left,xshift=2pt,yshift=-2pt] at (0,0.5) {$Q_2$};
\draw[thick] (0,0.5) -- (1,0.5);
\draw[fill] (0,0.5) circle [radius=2.0pt];
\draw[-triangle 90] (0.75,0.5) -- +(0.1,-0.0);
\node[above right,xshift=-2pt,yshift=-2pt] at (0,1.5) {$R_1$};
\draw[fill] (0,1.5) circle [radius=2.0pt];
\node[below right,xshift=-2pt,yshift=2pt] at (0,-0.5) {$R_2$};
\draw[fill] (0,-0.5) circle [radius=2.0pt];
\draw[thick] plot [smooth] coordinates {(0,2.5) (-0.5,2.0) (0,1.5) (1.5,0.5)
(0,-0.5) (-0.5,-0.0) (0,0.5)};
\node[below] at (0,-1) {(a)};
\end{tikzpicture}
~
\begin{tikzpicture}
\draw[dotted] (0,-0.8) -- (0,2.8);
\node[above,xshift=0pt] at (0,2.8) {$\ell[k]$};
\node[below right,xshift=-4pt,yshift=0pt] at (0,2.5) {$Q_1$};
\draw[thick] (0.5,2.5) -- (0,2.5);
\draw[fill] (0,2.5) circle [radius=2.0pt];
\draw[-triangle 90] (0.4,2.5) -- +(-0.1,-0.0);
\node[below right,xshift=-2pt,yshift=2pt] at (0,-0.5) {$Q_2$};
\draw[thick] (0,-0.5) -- (1,-0.5);
\draw[fill] (0,-0.5) circle [radius=2.0pt];
\draw[-triangle 90] (0.75,-0.5) -- +(0.1,-0.0);
\node[below right,xshift=-2pt,yshift=2pt] at (0,1.0) {$Q$};
\draw[fill] (0,1.0) circle [radius=2.0pt];
\node[above,xshift=6pt,yshift=-2pt] at (-1.0,1.0) {$P[k']$};
\draw[fill] (-1.0,1.0) circle [radius=2.0pt];
\draw[thick] plot [smooth] coordinates {(0,2.5) (-1.0,2.0) (-1.3,1.0) (-1.0,0.0) (0,-0.5)};
\draw[thick] plot [smooth] coordinates {(0.5,2.5) (1.0,1.8) (0,1.0) (-1.0,1.0)};
\node[below] at (0,-1) {(b)};
\end{tikzpicture}
~
\begin{tikzpicture}
\begin{scope}
\clip plot [smooth] coordinates {(0.5,2.5) (1.0,2.0) (0,1.5) (-1.0,1.0) (0,0.5)};
\fill[color=lightgray] (-2,-2) rectangle (0,2);
\end{scope}
\draw[dotted] (0,-0.8) -- (0,2.8);
\node[above,xshift=0pt] at (0,2.8) {$\ell[k]$};
\node[below right,xshift=-4pt,yshift=0pt] at (0,2.5) {$Q_1$};
\draw[thick] (0.5,2.5) -- (0,2.5);
\draw[fill] (0,2.5) circle [radius=2.0pt];
\draw[-triangle 90] (0.4,2.5) -- +(-0.1,-0.0);
\node[below right,xshift=-2pt,yshift=2pt] at (0,-0.5) {$Q_2$};
\draw[thick] (0,-0.5) -- (1,-0.5);
\draw[fill] (0,-0.5) circle [radius=2.0pt];
\draw[-triangle 90] (0.75,-0.5) -- +(0.1,-0.0);
\node[below right,xshift=-2pt,yshift=2pt] at (0,1.5) {$Q$};
\draw[fill] (0,1.5) circle [radius=2.0pt];
\node[below right,xshift=-2pt,yshift=2pt] at (0,0.5) {$Q'$};
\draw[fill] (0.0,0.5) circle [radius=2.0pt];
\draw[thick] (0,0.5) -- (1,0.5);
\draw[-triangle 90] (0.75,0.5) -- +(-0.1,-0.0);
\draw[thick] plot [smooth] coordinates {(0,2.5) (-1.0,2.0) (-1.3,1.0) (-1.0,0.0) (0,-0.5)};
\draw[thick] plot [smooth] coordinates {(0.5,2.5) (1.0,2.0) (0,1.5) (-1.0,1.0) (0,0.5)};
\node[below] at (0,-1) {(c)};
\end{tikzpicture}
\end{center}
\caption{(a) Regions (shaded) of an excursion at $\ell[k]$; due to 
additional crossings in $R_1$ and $R_2$, three more excursions exist, each with a smaller area.
(b) \& (c)
If truncation would introduce self-intersection, then either 
the excursion is filled, with some point $P[k']$ as in (b), or
there is an excursion with smaller area, illustrated by shading in (c).}
\label{fig:areas}
\end{figure}

The \emph{regions} of an excursion of $C$ between $Q_1$ and $Q_2$ at $\ell[k]$ 
are those that are enclosed by (subpaths of) $\sub_C(Q_1,Q_2)$
and (subsegments of) $\seg(Q_1,Q_2)$, as illustrated by
Figure~\ref{fig:areas}(a). The \emph{area} of the excursion is the 
surface area of all regions together.
We say an excursion is \emph{filled} if any of its regions
contains at least one point $P[k']$, some integer $k'$, 
otherwise it is said to be \emph{unfilled}.

We say $A$ and $B$ are in \emph{normal form} if no excursion can be truncated
without violating the four constraints or introducing a self-intersection.
Suppose $A$ and $B$ are in normal form, while one or more excursions 
remain. Let us first consider the unfilled excursions. Among them choose one 
that has the smallest area. By assumption, one of the four constraints must be
violated or a new self-intersection must be introduced, 
if we were to truncate that excursion. 
We will consider all relevant cases.

Each case will assume an unfilled excursion from the right (excursions from the left
are symmetric) of a path $C$ between $Q_1$ and $Q_2$
at $\ell[k]$. 
We may assume that $\sub_C(Q_1,Q_2)\cap \ell[k] = \{Q_1,Q_2\}$, as
additional crossings of $\ell[k]$ would mean that excursions exist with
smaller areas (cf.\ Figure~\ref{fig:areas}(a)), contrary to the assumptions.
Now assume truncation is blocked due to $Q\in \seg(Q_1,Q_2)\cap C$ such that 
$\dist_C(Q) < \dist_C(Q_1)$ (the case $\dist_C(Q_2) < \dist_C(Q)$ is
symmetric), as we need to preserve absence of self-intersection.
Suppose $Q$ is the only such point, so that $C$ crosses
$\seg(Q_1,Q_2)$ from left to right once without ever crossing it from right to
left, until $Q_1$ is reached. Then $C$ starts in the area of the excursion,
or in other words, the excursion is filled, contrary to the assumptions (cf.\
Figure~\ref{fig:areas}(b)).
Now suppose there are points $Q'$ and $Q$
where $C$ crosses $\seg(Q_1,Q_2)$ from right to left and from left to right,
respectively and $\dist_C(Q') < \dist_C(Q) < \dist_C(Q_1)$. 
If there are several choices, choose $Q'$ and $Q$ such that $\sub_C(Q',Q)\cap\ell[k] = \{Q',Q\}$.
This means the excursion between $Q'$ and $Q$ has an area smaller than the one between
$Q_1$ and $Q_2$, contrary to the assumptions (cf.\ Figure~\ref{fig:areas}(c)).

Note that excursions with zero area, that is, those that intersect with $\ell[k]$
without crossing over to the other side, can always be truncated. 
We can therefore further ignore non-crossing intersections.

\begin{figure}[t]
\begin{center}
\begin{tikzpicture}
\draw[dotted] (0,-0.8) -- (0,2.8);
\node[above,xshift=0pt] at (0,2.8) {$\ell[k]$};
\node[below right,xshift=-2pt,yshift=2pt] at (0,2.5) {$Q_1$};
\draw[thick] (1,2.5) -- (0,2.5);
\node[right] at (1,2.5) {$C$};
\draw[fill] (0,2.5) circle [radius=2.0pt];
\draw[-triangle 90] (0.75,2.5) -- +(-0.1,-0.0);
\node[below right,xshift=-2pt,yshift=2pt] at (0,-0.5) {$Q_2$};
\draw[thick] (0,-0.5) -- (1,-0.5);
\draw[fill] (0,-0.5) circle [radius=2.0pt];
\draw[-triangle 90] (0.75,-0.5) -- +(0.1,-0.0);
\draw[thick] (1,1.0) -- (-1,1.0);
\node[right] at (1,1.0) {$D$};
\node[below right,xshift=-2pt,yshift=2pt] at (0,1.0) {$Q$};
\draw[fill] (0,1.0) circle [radius=2.0pt];
\node[above,xshift=6pt,yshift=-2pt] at (-1.0,1.0) {$P[k']$};
\draw[fill] (-1.0,1.0) circle [radius=2.0pt];
\draw[thick] plot [smooth] coordinates {(0,2.5) (-1.0,2.0) (-1.3,1.0)
(-1.0,0.0) (0,-0.5)};
\node[below] at (0,-1) {(a)};
\end{tikzpicture}
~
\begin{tikzpicture}
\begin{scope}
\clip plot [smooth] coordinates {(0,1.5) (-1.0,1.0)
(0,0.5)};
\fill[color=lightgray] (-2,-2) rectangle (0,2);
\end{scope}
\draw[dotted] (0,-0.8) -- (0,2.8);
\node[above,xshift=0pt] at (0,2.8) {$\ell[k]$};
\node[below right,xshift=-2pt,yshift=2pt] at (0,2.5) {$Q_1$};
\draw[thick] (1,2.5) -- (0,2.5);
\node[right] at (1,2.5) {$C$};
\draw[fill] (0,2.5) circle [radius=2.0pt];
\draw[-triangle 90] (0.75,2.5) -- +(-0.1,-0.0);
\node[below right,xshift=-2pt,yshift=2pt] at (0,-0.5) {$Q_2$};
\draw[thick] (0,-0.5) -- (1,-0.5);
\draw[fill] (0,-0.5) circle [radius=2.0pt];
\draw[-triangle 90] (0.75,1.5) -- +(-0.1,-0.0);
\draw[thick] (1,1.5) -- (0,1.5);
\node[right] at (1,1.5) {$D$};
\node[below right,xshift=-2pt,yshift=2pt] at (0,1.5) {$Q$};
\draw[fill] (0,1.5) circle [radius=2.0pt];
\draw[-triangle 90] (0.75,0.5) -- +(0.1,-0.0);
\draw[thick] (1,0.5) -- (0,0.5);
\node[below right,xshift=-2pt,yshift=2pt] at (0,0.5) {$Q'$};
\draw[fill] (0.0,0.5) circle [radius=2.0pt];
\draw[thick] (0,0.5) -- (1,0.5);
\draw[-triangle 90] (0.75,-0.5) -- +(-0.1,-0.0);
\draw[thick] plot [smooth] coordinates {(0,2.5) (-1.0,2.0) (-1.3,1.0)
(-1.0,0.0) (0,-0.5)};
\draw[thick] plot [smooth] coordinates {(0,1.5) (-1.0,1.0) (0,0.5)};
\node[below] at (0,-1) {(b)};
\end{tikzpicture}
\end{center}
\caption{Truncating the excursion would introduce a violation of
constraint~(ii). The assumptions are contradicted in one of two ways.}
\label{fig:offend1}
\end{figure}
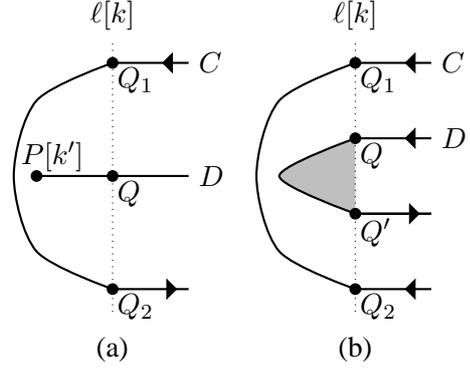

Now suppose a truncation would violate constraint~(ii), where $C=B[k'+1]$ and
$D=A[k']$ crosses $\seg(Q_1,Q_2)$. Then much as above, we may distinguish
two cases. In the first, $D$ has only one crossing of $\seg(Q_1,Q_2)$
in some point $Q$,
which means the excursion is filled with the starting or ending point of $D$, 
as in Figure~\ref{fig:offend1}(a).
In the second, $D$ has at least two consecutive crossings,
say in $Q$ and $Q'$, from right to left and from left to right, respectively,
which means the 
excursion between $Q$ and $Q'$ has smaller area than the one between
$Q_1$ and $Q_2$, illustrated by shading in Figure~\ref{fig:offend1}(b).
Both cases contradict the assumptions.
For $C=A[k']$ and $D=B[k'+1]$, the reasoning is symmetric.

\begin{figure}[t]
\begin{center}
\begin{tikzpicture}
\draw[dotted] (0,-0.8) -- (0,2.8);
\node[above,xshift=0pt] at (0,2.8) {$\ell[k]$};
\node[below right,xshift=-2pt,yshift=2pt] at (0,2.5) {$Q_1$};
\draw[thick] (1,2.5) -- (0,2.5);
\node[above,xshift=-2pt,yshift=-0pt] at (0.8,2.5) {$A[k']$};
\draw[fill] (0,2.5) circle [radius=2.0pt];
\draw[-triangle 90] (0.75,2.5) -- +(-0.1,-0.0);
\node[below right,xshift=-2pt,yshift=2pt] at (0,-0.5) {$Q_2$};
\draw[thick] (0,-0.5) -- (1,-0.5);
\draw[fill] (0,-0.5) circle [radius=2.0pt];
\draw[-triangle 90] (0.75,-0.5) -- +(0.1,-0.0);
\draw[thick] (1,1.0) -- (-1,1.0);
\node[above,xshift=-2pt,yshift=-1pt] at (0.9,1.0) {$A[k'-1]$};
\node[below right,xshift=-2pt,yshift=2pt] at (0,1.0) {$Q$};
\draw[fill] (0,1.0) circle [radius=2.0pt];
\draw[-triangle 90] (0.75,1.0) -- +(-0.1,-0.0);
\node[above,xshift=6pt,yshift=-2pt] at (-1.0,1.0) {$P[k']$};
\draw[fill] (-1.0,1.0) circle [radius=2.0pt];
\draw[thick] plot [smooth] coordinates {(0,2.5) (-1.0,2.0) (-1.3,1.0)
(-1.0,0.0) (0,-0.5)};
\node[below] at (0,-1) {(a)};
\end{tikzpicture}
~
\begin{tikzpicture}
\begin{scope}
\clip plot [smooth] coordinates {(0,2.5) (-1.0,2.0) (-1.5,0.8)
(-1.0,-0.5) (-0.5,-0.5) (0,0)};
\fill[color=lightgray] (-2,-2) rectangle (-1,2);
\end{scope}
\draw[dotted] (0,-0.8) -- (0,2.8);
\node[above,xshift=0pt] at (0,2.8) {$\ell[k]$};
\draw[dotted] (-1,-0.8) -- (-1,2.8);
\node[above,xshift=0pt] at (-1,2.8) {$\ell[k{-}1]$};
\node[below right,xshift=-2pt,yshift=2pt] at (0,2.5) {$Q_1$};
\draw[thick] (1,2.5) -- (0,2.5);
\node[above,xshift=-2pt,yshift=-0pt] at (0.8,2.5) {$A[k']$};
\draw[fill] (0,2.5) circle [radius=2.0pt];
\draw[-triangle 90] (0.75,2.5) -- +(-0.1,-0.0);
\node[below right,xshift=-2pt,yshift=2pt] at (0,0.0) {$Q_2$};
\draw[thick] (0,0.0) -- (1,0.0);
\draw[fill] (0,0.0) circle [radius=2.0pt];
\draw[-triangle 90] (0.75,0.0) -- +(0.1,-0.0);
\draw[thick] (1,1.0) -- (0,1.0);
\node[above,xshift=-2pt,yshift=-1pt] at (0.9,1.0) {$A[k'-1]$};
\node[below right,xshift=-2pt,yshift=2pt] at (0,1.0) {$Q$};
\draw[-triangle 90] (0.75,1.0) -- +(0.1,-0.0);
\draw[fill] (0,1.0) circle [radius=2.0pt];
\node[below right,xshift=-2pt,yshift=5pt] at (-1.0,0.0) {$Q_2'$};
\draw[fill] (-1.0,0.0) circle [radius=2.0pt];
\draw[thick] plot [smooth] coordinates {(0,2.5) (-1.0,2.0) (-1.5,0.8)
(-1.0,-0.5) (-0.5,-0.5) (0,0)};
\draw[thick] plot [smooth] coordinates {(0,1.0) (-0.5,0.7) (-1.0,0.0)};
\draw[fill] (-1,-0.5) circle [radius=2.0pt];
\node[below right,xshift=-2pt,yshift=2pt] at (-1,2.0) {$R_1$};
\node[below left,xshift=2pt,yshift=2pt] at (-1,-0.5) {$R_2$};
\draw[fill] (-1,2.0) circle [radius=2.0pt];
\node[below] at (0,-1) {(b)};
\end{tikzpicture}
\end{center}
\caption{Truncating the excursion would introduce a violation of
constraint~(iii), where $\dist_{A[k'-1]}(Q) > \dist_{A[k']}(Q_2)$.
The assumptions are contradicted in one of three ways, the first as
in Figure~\ref{fig:offend1}(b), and the second and third as in
(a) and (b) above.}
\label{fig:offend2a}
\end{figure}
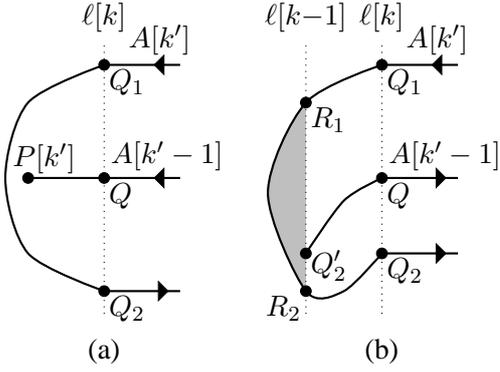

Next, suppose a truncation would violate constraint~(iii),
where $C=A[k']$ and $A[k'-1]$ crosses $\seg(Q_1,Q_2)$ in $Q$, while
$\dist_{A[k'-1]}(Q) > \dist_{A[k']}(Q_2)$. If the crossing in $Q$ is
from right to left, and there is an immediately next crossing in $Q'$ from
left to right, then we have the same situation as in
Figure~\ref{fig:offend1}(b), involving an excursion with smaller area, contradicting the assumptions.
If the crossing in $Q$ is the only one,
and it is from right to left,
then we can use the fact that $\sub_{A[k']}(Q_1,Q_2)\cap \sub_{A[k'-1]}(Q,P[k'])=\emptyset$,
as we assume the four constraints as yet hold. This means $P[k']$ must be
contained in the area of the excursion, as illustrated in
Figure~\ref{fig:offend2a}(a), contradicting the assumption that the
excursion is unfilled.
If the crossing in $Q$ is the only one,
and it is from left to right, then we can use the fact that
$\sub_{A[k']}(Q_1,Q_2)\cap \sub_{A[k'-1]}(Q_2',Q)=\emptyset$, for the unique
$Q_2'\in A[k'-1]\cap \ell[k-1]$ such that $\dist_{A[k'-1]}(Q_2') = \dist_{A[k']}(Q_2)$. This means
the excursion contains $Q_2'$,
which implies there is another unfilled excursion between points $R_1,R_2 \in
A[k'] \cap \ell[k-1]$ with smaller area, as shaded in Figure~\ref{fig:offend2a}(b),
contrary to the assumptions.


Suppose a truncation would violate constraint~(iii),
where $C=A[k']$ and $A[k'+1]$ crosses $\seg(Q_1,Q_2)$ in $Q$, while
$\dist_{A[k']}(Q_1) > \dist_{A[k'+1]}(Q)$. The reasoning is now largely
symmetric to the above, with the direction of the crossing reversed,
except that the case analogous to
Figure~\ref{fig:offend2a}(b) is immediately excluded, as $Q_2'$ cannot be
both to the left and to the right of $\ell[k]$.
Constraint~(iv) is symmetric to constraint~(iii). All possible cases
have been shown to lead to contradictions, and therefore we conclude that 
there are no unfilled excursions if $A$ and $B$ are in normal form.

We now show that there cannot be any filled excursions either.
For this, assume that $A[k']$ has a filled excursion between $Q_1$ and $Q_2$
at $\ell[k]$ from the right.
This means $A[k'-1]$ has an identically shaped, filled excursion at $\ell[k-1]$ from the right,
between corresponding points $Q'_1$ and $Q'_2$.
Let us consider how path $A[k']$ proceeds after reaching $Q_2$. 
There are only three possibilities:
\begin{itemize}
\item it ends in $P[k+1]$, with $k'=k$, without further 
crossings of $\ell[k]$ or $\ell[k+1]$;
\item it next crosses $\ell[k]$ leftward; or
\item it next crosses $\ell[k+1]$ in some point $Q_3$.
\end{itemize}
The first of these can be excluded, in the light of 
$\dist_{A[k'-1]}(Q) \geq \dist_{A[k']}(Q_2)$ for each
$Q\in\sub_{A[k'-1]}(Q'_2,P[k'])$.
Due to constraint~(iii) therefore, this subpath of $A[k'-1]$ cannot intersect
with the
excursion of $A[k']$ to reach $P[k]$, and therefore $A[k']$ cannot reach $P[k+1]$. 
The second possibility is also excluded, as this would
imply the existence of an unfilled excursion. 
For the remaining possibility, $Q_3\in A[k']\cap \ell[k+1]$ may be lower down than $Q_2$ (in the now familiar
view of the points $P[0],P[1],\ldots$ being drawn from left to right along a
horizontal line), or it may be higher up than $Q_1$.
These two cases are drawn in Figures~\ref{fig:filled1} and~\ref{fig:filled2}.
The choice of $Q_3$ also determines a corresponding $Q'_3\in A[k'-1]\cap \ell[k]$.

We now consider how $A[k']$ continues after $Q_3$ in the case of Figure~\ref{fig:filled1}.
If it next crosses $\ell[k+1]$ leftward, this would imply the existence of
an unfilled excursion. Further,
$\dist_{A[k'-1]}(Q) \geq \dist_{A[k']}(Q_3)$ for each
$Q\in\sub_{A[k'-1]}(Q'_3,P[k'])$.
Due to constraint~(iii) therefore, this subpath of $A[k'-1]$ cannot intersect
with $\sub_{A[k']}(Q_2,Q_3)$, above which lies $P[k+1]$. Therefore, 
$A[k']$ must cross $\ell[k+2]$ in some $Q_4$, which is lower down than $Q_3$.
This continues ad infinitum, and $A[k']$ will never reach its supposed end point $P[k'+1]$. 
The reasoning for Figure~\ref{fig:filled2} is similar.
\begin{figure}[t]
\begin{center}
\begin{tikzpicture}[scale=0.98]
\begin{scope}[shift={(-2,0)}]
\draw[dotted] (0,-0.3) -- (0,2.3);
\node[above,xshift=0pt] at (0,2.2) {$\ell[k{-}1]$};
\draw[fill] (0,2.0) circle [radius=2.0pt];
\node[below right,xshift=-2pt,yshift=2pt] at (0,2.0) {$Q'_1$};
\draw[fill] (0,1.0) circle [radius=2.0pt];
\draw[thick,dashed] plot [smooth] coordinates {(0,2.0) (-1.0,2.0) (-1.3,1.5)
(-1.0,1.0) (0,1.0)};
\node[below right,xshift=-2pt,yshift=2pt] at (0,1.0) {$Q'_2$};
\draw[fill] (-1.0,1.5) circle [radius=2.0pt];
\node[above right,xshift=-7pt,yshift=-2pt,font=\small] at (-1.0,1.5) {$P[k{-}1]$};
\draw[thick,dashed] plot [smooth] coordinates {(0,1.0) (0.5,1.0) (1.0,0.5) (2,0.5)};
\draw[fill] (2,0.5) circle [radius=2.0pt];
\node[below right,xshift=-2pt,yshift=2pt] at (2,0.5) {$Q'_3$};
\draw[thick,dashed] plot [smooth] coordinates {(2,0.5) (2.5,0.5) (3.0,0.0) (4,0.0)};
\draw[fill] (4,0.0) circle [radius=2.0pt];
\node[below right,xshift=-2pt,yshift=2pt] at (4,0.0) {$Q'_4$};
\end{scope}
\draw[dotted] (0,-0.3) -- (0,2.3);
\node[above,xshift=0pt] at (0,2.2) {$\ell[k]$};
\draw[-triangle 90] (-0.20,2.0) -- +(-0.1,-0.0);
\draw[fill] (0,2.0) circle [radius=2.0pt];
\node[below right,xshift=-2pt,yshift=2pt] at (0,2.0) {$Q_1$};
\draw[fill] (0,1.0) circle [radius=2.0pt];
\draw[thick] (0,2.0) -- plot [smooth] coordinates {(-0.20,2.0) (-1.0,2.0) (-1.3,1.5) (-1.0,1.0) (0,1.0)};
\node[below right,xshift=-2pt,yshift=2pt] at (0,1.0) {$Q_2$};
\draw[fill] (-1.0,1.5) circle [radius=2.0pt];
\node[above right,xshift=-7pt,yshift=-2pt,font=\small] at (-1.0,1.5) {$P[k]$};
\draw[thick] plot [smooth] coordinates {(0,1.0) (0.5,1.0) (1.0,0.5) (2,0.5)};
\draw[fill] (2,0.5) circle [radius=2.0pt];
\node[below right,xshift=-2pt,yshift=2pt] at (2,0.5) {$Q_3$};
\draw[thick] plot [smooth] coordinates {(2,0.5) (2.5,0.5) (3.0,0.0) (4,0.0)};
\draw[fill] (4,0.0) circle [radius=2.0pt];
\node[below right,xshift=-2pt,yshift=2pt] at (4,0.0) {$Q_4$};
\begin{scope}[shift={(2,0)}]
\draw[dotted] (0,-0.3) -- (0,2.3);
\node[above,xshift=0pt] at (0,2.2) {$\ell[k{+}1]$};
\draw[fill] (-1.0,1.5) circle [radius=2.0pt];
\node[above right,xshift=-7pt,yshift=-2pt,font=\small] at (-1.0,1.5) {$P[k{+}1]$};
\end{scope}
\begin{scope}[shift={(4,0)}]
\draw[dotted] (0,-0.3) -- (0,2.3);
\node[above,xshift=0pt] at (0,2.2) {$\ell[k{+}2]$};
\draw[fill] (-1.0,1.5) circle [radius=2.0pt];
\node[above right,xshift=-7pt,yshift=-2pt,font=\small] at (-1.0,1.5) {$P[k{+}2]$};
\end{scope}
\end{tikzpicture}
\end{center}
\caption{Continuing the (solid) path $A[k']$ after a filled excursion,
restricted by the (dashed) path $A[k'-1]$, in the light of constraint~(iii).}
\label{fig:filled1}
\begin{center}
\begin{tikzpicture}[scale=0.98]
\begin{scope}[shift={(-2,0)}]
\draw[dotted] (0,0.4) -- (0,3.3);
\node[above,xshift=0pt] at (0,3.2) {$\ell[k{-}1]$};
\draw[fill] (0,2.0) circle [radius=2.0pt];
\node[below right,xshift=-2pt,yshift=2pt] at (0,2.0) {$Q'_1$};
\draw[fill] (0,1.0) circle [radius=2.0pt];
\draw[thick,dashed] plot [smooth] coordinates {(0,2.0) (-1.0,2.0) (-1.3,1.5)
(-1.0,1.0) (0,1.0)};
\node[below right,xshift=-2pt,yshift=2pt] at (0,1.0) {$Q'_2$};
\draw[fill] (-1.0,1.5) circle [radius=2.0pt];
\node[above right,xshift=-7pt,yshift=-2pt,font=\small] at (-1.0,1.5) {$P[k{-}1]$};
\draw[thick,dashed] plot [smooth] coordinates {(0,1.0) (0.5,1.2) (0.8,2.2) (2,2.5)};
\draw[fill] (2,2.5) circle [radius=2.0pt];
\node[below right,xshift=-2pt,yshift=3pt] at (2,2.5) {$Q'_3$};
\draw[thick,dashed] plot [smooth] coordinates {(2,2.5) (3,2.7) (4,3.0)};
\draw[fill] (4,3.0) circle [radius=2.0pt];
\node[below right,xshift=-2pt,yshift=5pt] at (4,3.0) {$Q'_4$};
\end{scope}
\draw[dotted] (0,0.4) -- (0,3.3);
\node[above,xshift=0pt] at (0,3.2) {$\ell[k]$};
\draw[-triangle 90] (-0.2,2.0) -- +(-0.1,-0.0);
\draw[fill] (0,2.0) circle [radius=2.0pt];
\node[below right,xshift=-2pt,yshift=2pt] at (0,2.0) {$Q_1$};
\draw[fill] (0,1.0) circle [radius=2.0pt];
\draw[thick] (0,2.0) -- plot [smooth] coordinates {(-0.2,2.0) (-1.0,2.0) (-1.3,1.5) (-1.0,1.0) (0,1.0)};
\node[below right,xshift=-2pt,yshift=2pt] at (0,1.0) {$Q_2$};
\draw[fill] (-1.0,1.5) circle [radius=2.0pt];
\node[above right,xshift=-7pt,yshift=-2pt,font=\small] at (-1.0,1.5) {$P[k]$};
\draw[thick] plot [smooth] coordinates {(0,1.0) (0.5,1.2) (0.8,2.2) (2,2.5)};
\draw[fill] (2,2.5) circle [radius=2.0pt];
\node[below right,xshift=-2pt,yshift=2pt] at (2,2.5) {$Q_3$};
\draw[thick] plot [smooth] coordinates {(2,2.5) (3,2.7) (4,3.0)};
\draw[fill] (4,3.0) circle [radius=2.0pt];
\node[below right,xshift=-2pt,yshift=2pt] at (4,3.0) {$Q_4$};
\begin{scope}[shift={(2,0)}]
\draw[dotted] (0,0.4) -- (0,3.3);
\node[above,xshift=0pt] at (0,3.2) {$\ell[k{+}1]$};
\draw[fill] (-1.0,1.5) circle [radius=2.0pt];
\node[above right,xshift=-7pt,yshift=-2pt,font=\small] at (-1.0,1.5) {$P[k{+}1]$};
\end{scope}
\begin{scope}[shift={(4,0)}]
\draw[dotted] (0,0.4) -- (0,3.3);
\node[above,xshift=0pt] at (0,3.2) {$\ell[k{+}2]$};
\draw[fill] (-1.0,1.5) circle [radius=2.0pt];
\node[above right,xshift=-7pt,yshift=-2pt,font=\small] at (-1.0,1.5) {$P[k{+}2]$};
\end{scope}
\end{tikzpicture}
\end{center}
\caption{As in Figure~\ref{fig:filled1} but $Q_3$ is chosen to be higher up than
$Q_1$.}
\label{fig:filled2}
\end{figure}

Filled excursions from the left are symmetric, but instead of investigating the path
after $Q_2$, we must investigate the path \emph{before} $Q_1$. The case of $B$ is
symmetric to that of $A$. We may now conclude no filled excursions exist.

\section{The final contradiction}
\label{sec:final}

We have established that after $A$ and $B$ have been brought into normal form,
there can be no remaining excursions. This means that $A[0]$ crosses $\ell[0]$
exactly once, in some point $R_A$, and $B[0]$ crosses $\ell[-1]$ exactly once,
in some point $L_B$. Further, let $L_A$ be the
unique point where $A[-1]$ crosses $\ell[-1]$ and $R_B$ the unique point where
$B[1]$ crosses $\ell[0]$.

\begin{figure}[t]
\begin{center}
\begin{tikzpicture}[scale=0.45]
\begin{scope}
\clip plot (0,0) -- plot [smooth] coordinates {
(-1,-1) (1.5,-3) (2.3,0) (1,3)
(3.0,3.0)} -- (3.5,3.0) -- (3.5,3.5) -- (-3.5,3.5);
\clip (0,0) -- plot [smooth] coordinates {
(1,-1) (2,-1.5) (3,0) (2,2) (-1,2.5)
(-3.0,2.0)} -- (-3.5,2.0) -- (-3.5,3) -- (3.5,3) -- (3.5,-3) -- (-3.5,-3) --
(-3.5,0);
\clip (-3,-3) rectangle (3,0);
\fill[color=gray] (-3.5,-3) rectangle (3.5,3.5);
\end{scope}
\begin{scope}
\clip plot (0,0) -- plot [smooth] coordinates {
(-1,-1) (1.5,-3) (2.3,0) (1,3)
(3.0,3.0)} -- (3.5,3.0) -- (3.5,-3.5) -- (-3.5,-3.5) -- (-3.5,0);
\clip (0,0) -- plot [smooth] coordinates {
(1,-1) (2,-1.5) (3,0) (2,2) (-1,2.5)
(-3.0,2.0)} -- (-3.5,2.0) -- (-3.5,0);
\fill[color=gray] (-3.5,-3) rectangle (3.5,3.5);
\end{scope}
\begin{scope}
\clip plot (0,0) -- plot [smooth] coordinates {
(-1,-1) (1.5,-3) (2.3,0) (1,3)
(3.0,3.0)} -- (3.5,3.0) -- (3.5,5) -- (-3.5,5) -- (-3.5,0);
\clip (0,0) -- plot [smooth] coordinates {
(1,-1) (2,-1.5) (3,0) (2,2) (-1,2.5)
(-3.0,2.0)} -- (-3.5,2.0) -- (-3.5,5) -- (3.5,5) -- (3.5,-3) -- (-3.5,-3) --
(-3.5,0);
\clip (-3.5,0) rectangle (3.5,5);
\fill[color=lightgray] (-3.5,-3) rectangle (3.5,5);
\end{scope}
\draw[dashed] (3.5,-4) -- (3.5,5);
\draw[dashed] (-3.5,-4) -- (-3.5,5);
\node[above,yshift=-1pt,xshift=-0pt] at (-3.5,5.0) {$\ell[-1]$};
\node[above,yshift=-1pt,xshift=-0pt] at (3.5,5.0) {$\ell[0]$};
\node[above,xshift=-4pt,yshift=3.0pt] at (-7,-0.0) {$P[-1]$};
\draw[fill] (-7,0) circle [radius=4.0pt];
\node[right,xshift=-0.0pt,yshift=3pt] at (0,0.0) {$P[0]$};
\draw[fill] (0,0) circle [radius=4.0pt];
\node[above,xshift=4pt,yshift=2.0pt] at (7,-0.0) {$P[1]$};
\draw[fill] (7,0) circle [radius=4.0pt];
\node[above,xshift=-0pt,yshift=1pt] at (1,3) {$A[0]$};
\draw [thick] (0,0) -- plot [smooth] coordinates {
(-1,-1) (1.5,-3) (2.3,0) (1,3)
(3.0,3.0)} -- (3.5,3.0);
\node[above,xshift=-0pt,yshift=-1pt] at (-1,2.5) {$B[0]$};
\draw [thick] (0,0) -- plot [smooth] coordinates {
(1,-1) (2,-1.5) (3,0) (2,2) (-1,2.5)
(-3.0,2.0)} -- (-3.5,2.0);
\node[below,xshift=-2pt,yshift=-8pt] at (-1.5,-2) {$B[1]$};
\draw [dashed] (0,0) -- plot [smooth] coordinates {
(-0.5,0.5) (-2,1.5) (-1.5,-2) (0,-3.3) (3.0,-3.3)} -- (3.5,-3.3);
\node[left,xshift=-4pt,yshift=-15pt,fill=white,rounded corners=2pt,inner sep=2pt] at (-2.0,0.7) {$A[-1]$};
\draw [dashed] (0,0) -- plot [smooth] coordinates {
(-0.5,0.5) (-2.0,0.7) (-2.5,-1.5) (-3.0,-2)} -- (-3.5,-2);
\draw [dotted] (-7,-0.0) -- (-3.5,2.0);
\draw [dotted] (-7,-0.0) -- (-3.5,-2.0);
\draw [dotted] (7,-0.0) -- (3.5,3.0);
\draw [dotted] (7,-0.0) -- (3.5,-3.3);
\draw[fill] (-3.5,-2.0) circle [radius=4.0pt];
\draw[fill] (3.5,-3.3) circle [radius=4.0pt];
\draw[fill] (-3.5,2.0) circle [radius=4.0pt];
\draw[fill] (3.5,3.0) circle [radius=4.0pt];
\node[left,xshift=-0pt,yshift=3pt] at (-3.5,2.0) {$L_B$};
\node[left,xshift=-0pt,yshift=-1pt] at (-3.5,-2.0) {$L_A$};
\node[right,xshift=-0pt,yshift=1pt] at (3.5,3.0) {$R_A$};
\node[right,xshift=-0pt,yshift=-1pt] at (3.5,-3.3) {$R_B$};
\draw[-triangle 90] (-0.5,-0.5) -- +(-0.1,-0.1);
\draw[-triangle 90] (0.5,-0.5) -- +(0.1,-0.1);
\draw[-triangle 90] (3.2,3.0) -- +(0.1,0.0);
\draw[-triangle 90] (-3.2,2.0) -- +(-0.1,0.0);
\draw[-triangle 90] (3.0,-3.3) -- +(-0.1,0.0);
\draw[-triangle 90] (-3.0,-2) -- +(0.1,0.0);
\end{tikzpicture}
\end{center}
\caption{The region between $\ell[-1]$ and $\ell[0]$ is divided by
$A[0]$ and
$B[0]$ into a `top' region (lightly shaded),
a `bottom' region (white),
and areas enclosed by intersections of $A[0]$ and
$B[0]$ (darkly shaded). Here $A[-1]$ and $B[1]$ are both in the `bottom'
region.}
\label{fig:partition}
\end{figure}
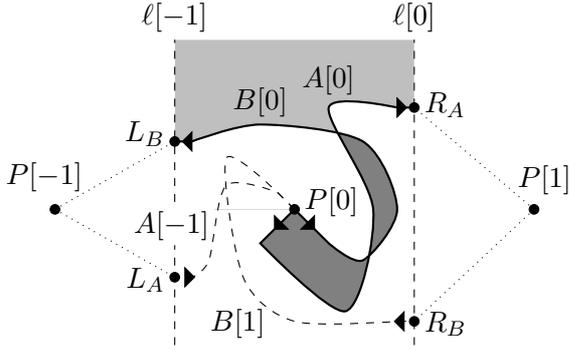

The region of the plane between $\ell[-1]$ and $\ell[0]$ can now be partitioned into
a `top' region, a `bottom' region, and zero or more enclosed regions. The `top'
region consists of those points that are
reachable from any point between $\ell[-1]$ and $\ell[0]$ arbitrarily far
above any point of $A[0]$ and $B[0]$, without intersecting with $A[0]$, $B[0]$,
$\ell[-1]$ or $\ell[0]$. This is the lightly shaded region in
Figure~\ref{fig:partition}. The `bottom' region is similarly defined,
in terms of reachability 
from any point between $\ell[-1]$ and $\ell[0]$ arbitrarily far \emph{below} any point of $A[0]$ and $B[0]$.
The zero or more enclosed regions 
stem from possible intersections of $A[0]$ and $B[0]$;
the two such enclosed regions in Figure~\ref{fig:partition} are darkly
shaded. Note that the four constraints do \emph{not} preclude intersections of
$A[0]$ and $B[0]$.

However, constraint~(ii) implies that, between $\ell[-1]$ and $\ell[0]$,
$A[0]$ and $B[1]$ do not intersect other than in $P[0]$, 
and similarly,
$A[-1]$ and $B[0]$ do not intersect other than in $P[0]$.
Moreover, for any $Q\in \sub_{A[-1]}(L_A,P[0])$ and
any $Q'\in \sub_{A[0]}(P[0],R_A)$ we have
$\dist_{A[-1]}(Q) \geq \dist_{A[0]}(Q')$. By constraint~(iii) this 
means
$A[-1]$ and $A[0]$ do not intersect other than in $P[0]$.
Similarly,
$B[1]$ and $B[0]$ do not intersect other than in $P[0]$.

The angles in $P[0]$ between $A[0]$, $B[0]$,
$A[-1]$ and $B[1]$ are multiples of $90\degree$.
Because of constraint~(i), which excludes a $180\degree$ angle between 
$A[0]$ and $B[0]$, it follows that either $\sub_{A[-1]}(L_A,P[0])$ and
$\sub_{B[1]}(R_B,P[0])$ both lie entirely 
in the `top' region, or both lie entirely in the `bottom' region. 
The latter case 
is illustrated in Figure~\ref{fig:partition}. In the former case, 
$L_A$ and $R_B$ are above $L_B$ and $R_A$, respectively, 
and in the latter case
$L_A$ and $R_B$ are below $L_B$ and $R_A$. This is impossible,
as $L_A$ and $R_A$ should be at the same height, these being corresponding
points of $A[-1]$ and $A[0]$, which have the same shape,
and similarly $L_B$ and $R_B$ should be at the same height.

This contradiction now leads back to the very beginning of our proof, and 
implies that the four constraints cannot all be true, and therefore that
at least one rule is always applicable to allow use of the inductive
hypothesis, and therefore that $\mygram$ generates $\myo{2}$.

\section{Conclusions and outlook}

We have presented a new proof that $\myo{2}$ is generated by a MCFG. 
It has at least superficial elements in common with the proof
by \newcite{SA15}. Both proofs use essentially the same MCFG, both are
geometric in nature, and both involve a continuous view of paths next to a
discrete view. The major difference lies in the approach to tackling the
myriad ways in which the paths can wind around each other and themselves. 
In the case of \newcite{SA15}, the key concept is that of the complex exponential
function, which seems to restrict the proof technique to two-dimensional geometry. 
In our case, the key concepts are excursions and truncation thereof. 

At this time, no proof is within reach that generalizes the result to $\myo{3}$.
This will require at the very least a suitable definition of `excursion'
in three dimensions,
which is likely to be more involved than in the two-dimensional case.
There are good reasons to believe however that if such a
definition can be found, some key elements of the existing proof may carry over. 
Instead of a binary predicate $R$ for our grammar $\mygram$, the appropriate
grammar for $\myo{3}$ would have a ternary predicate. Where our reasoning 
had two paths $A$ and $B$, the three dimensional case would involve 
three paths $A$, $B$ and $C$. In place of the reasoning illustrated in 
Figure~\ref{fig:partition}, where
$A$ and $B$ lead away from $P[0]$ to $P[-1]$ and $P[1]$, to partition the
plane into a `top' region and a `bottom' region, we would now have 
$A$ and $B$ and $C$ leading away from a central point, and prefixes of these
three paths would combine to partition the three-dimensional space into a 
`top' region and a `bottom' region. Copies of $A$ and $B$ and $C$
leading \emph{towards} the central point would then all
fall in the `top' region or all in the `bottom' region, 
potentially leading to a contradiction.  
Further details 
are beyond the scope of the present paper.

\section*{Acknowledgements}

This work came out of correspondence with Giorgio Satta. Gratefully
acknowledged are also fruitful discussions with Sylvain Salvati, Vinodh Rajan, 
and Markus Pfeiffer.

\bibliographystyle{acl2016}
\bibliography{/cs/home/mjn/bib/refs}

\end{document}